\documentclass[twocolumn,twoside,slac_two]{revtex4}
\usepackage{graphicx}
\usepackage{fancyhdr}
\begin{document}

\title{Using the SED to locate the {\boldmath $\gamma$}-ray emission site of powerful blazars}


\author{M. Georganopoulos}
\affiliation{ Department of Physics, University of Maryland Baltimore County, Baltimore MD 21250, USA}
\author{E. T. Meyer, G. Fossati}
\affiliation{Department of Physics and Astronomy, Rice University, Houston, TX 77005, USA}

\begin{abstract}
The location of the Gamma-ray emission of powerful blazars is a matter of active
debate. Is the location within the UV emitting sub-pc scale broad line region,
or farther out at pc scales where the molecular torus IR emission dominates?
We present a diagnostic that connects three observables, the synchrotron and
external Compton peak frequencies and the Compton dominance (the ratio of Compton
to synchrotron luminosity) to the seed photon energy and energy density. We
discuss encouraging preliminary results and discuss how  to use our diagnostic to
understand the location of the Gamma-ray emission as a function of source power
through the use of multiwavelength observations.
% including WISE, Planck, BAT, SWIFT and LAT.
\end{abstract}

\maketitle

\thispagestyle{fancy}


\section{Introduction}

A central question that is been debated in the
 {\sl Fermi} era regards the location of the blazar $\gamma$-ray
 emission site: is the $\gamma$ - ray emission of powerful blazars
 produced inside the sub-pc size broad line region (BLR) or further
 out at scales of $\sim 1$ - few pc (fig. 1) where the IR photon field
 of the dusty molecular torus (MT) dominates over that of the UV field
 of the BLR? In the first case the $\gamma$ - ray emission is most
 probably external Compton (EC) scattering of the $\sim$ 10 eV BLR
 photons \cite{sikora94}, while in the second the seed photons for the
 EC emission are the $\sim 0.1 $ eV photons emitted by the dust
 \cite{cleary07} in the MT \cite{blazejowski00}.  The issue of the
 energy dissipation location is connected to the jet formation and
 collimation process \cite{vlahakis04, marscher08} and, as we propose,
 can be approached through a SED diagnostic we introduce.

\begin{figure}[t]
\centerline{
\includegraphics [width=3.7in]{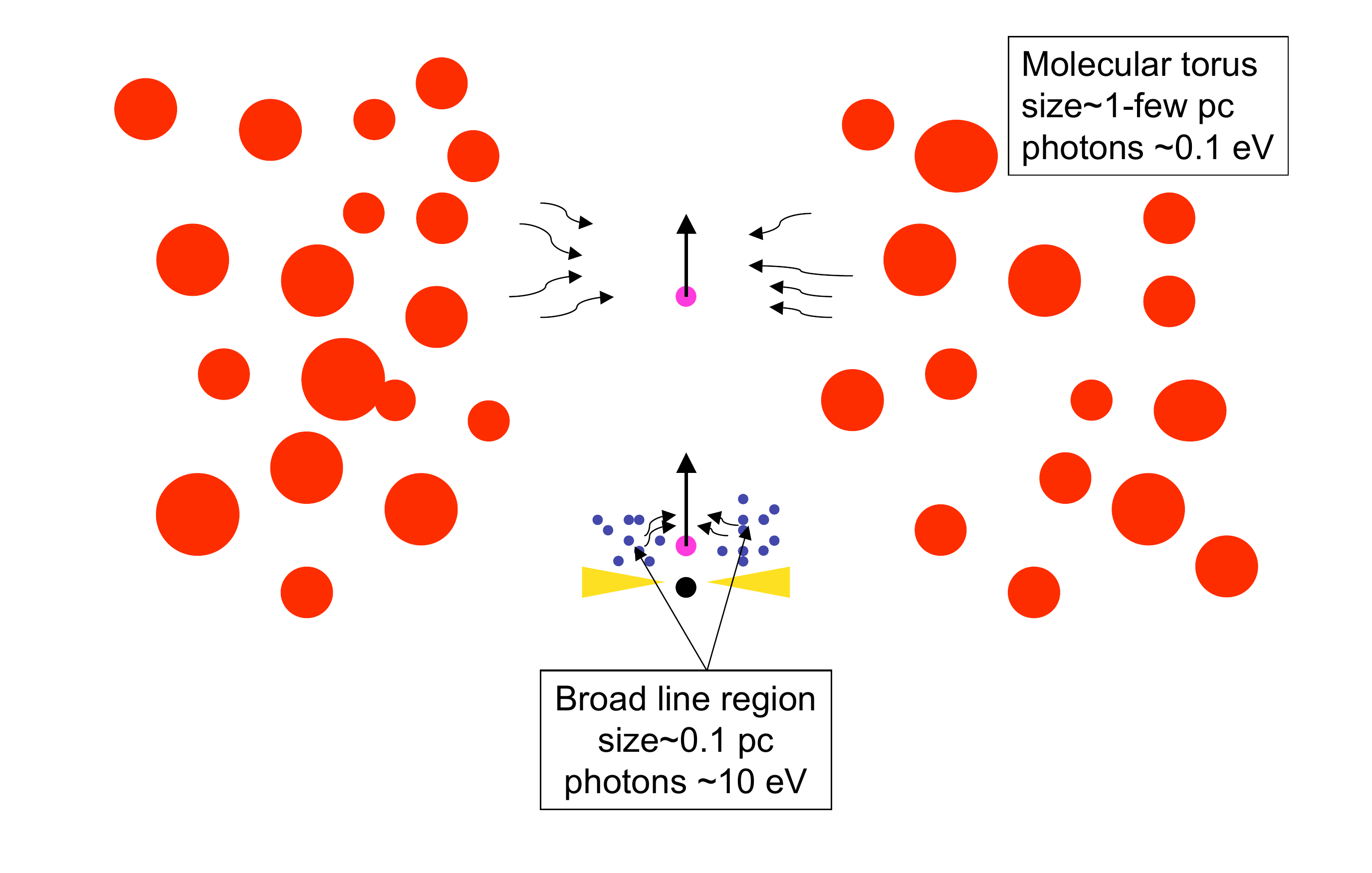}}
\caption{  Is the blazar GeV emission produced inside the
  sub-pc broad line region or further out at distances $\sim 1$ - few
  pc, comparable to the size of the molecular torus?  This is an ongoing debate
  with strong cases presented by both camps. Here we introduce a
  diagnostic based on the spectral energy distribution (SED) and we
  show how it can be used  to locate the blazar emission site.}
\label{powerfigure}
\end{figure}

\subsection{ The near camp: Few hour GeV variability  puts the  blazar inside the BLR.}
{\sl Fermi} has detected flares with decay time down to $\sim 3$
hours, comparable to the telescope sky scanning period. Such
variations have been seen in 3C 454.3 \cite{tavecchio10, ackermann10},
PKS 1454-354 \cite{abdo09}, 3C 273 \cite{abdo10b}, PKS 1502+106
\cite{abdo10a}, PKS B1222+216 \cite{foschini11}.  The short
variability timescale has been used to argue that the emission is
produced within the BLR (e.g. \cite{tavecchio10}): For a jet with
$\Gamma=\delta=10$, where $\Gamma$ is the bulk Lorentz factor of the
flow and $\delta$ the corresponding Doppler factor, variability times
of $t_{var}=10^4 $ s ($\sim$ 3 hours), correspond to a maximum source
size of $r=c t_{var} \delta=3\times 10^{15}$ cm.
Assuming a jet half-angle $\theta_{jet}=0.17/\Gamma$ \cite{jorstad05}, an upper limit on  the distance of the blazar from the central engine is $R=r/\theta_{jet}=r \Gamma/0.17= 3.5 \times 10^{17}$ cm. This distance is  comparable to  the BLR size $R_{BLR}\approx 1-3 \times 10^{17}$  cm\cite{kaspi07} and, therefore,  it is plausible that the GeV emission is produced inside the BLR.
% Assuming a jet opening angle $\theta_{jet}=1/\Gamma$, an upper limit on the distance of the blazar from the central engine is $R=r \Gamma= 3\times 10^{16}$ cm. This distance is smaller than the BLR size $R_{BLR}\approx 1-3 \times 10^{17}$ cm \cite{kaspi07,bentz09} and, therefore, the GeV emission must be produced inside the BLR. 
Essentially this argument is
based on the assumption that the entire cross section of the jet is
emitting. If however one allows smaller parts of the jet to produce
the $\gamma$-ray emission (e.g. \cite{giannios09}) such short
variations can take place further out.

\subsection { The far camp: Optical/VLBI polarization and GeV/X-ray data put
  the blazar few pc from the central engine.} The sub-pc scale energy
dissipation is challenged by observations that put the emission at few
pc distance from the black hole
(e.g. \cite{marscher08,abdo10c,marscher10a,jorstad10,agudo11}). In
several cases, optical polarimetry during an optical - $\gamma$-ray
flare showed a polarization behavior similar to that observed in
simultaneous high frequency VLBI coming from the 43 GHz core, several
pc away from the central engine. Because of the similar optical and
VLBI polarization behavior, the optical emission is constrained to
emerge from the VLBI core and because the optical and $\gamma$-ray
variations are seen to be simultaneous, the $\gamma$-ray emission is
also constrained to emerge from the VLBI core at a distance of few to
several pc from the black hole. Additional arguments placing some of
the $\gamma$-ray flares at $\sim 10$ pc have been advanced for 3C
454.3 \cite{sikora08,sikora09}. These arguments are based on the
similar behavior of optical and millimeter light curves and explain
the GeV emission as EC scattering of photons emitted from the dust of
the MT.  Although at distances of $\sim 10$ pc from the central
engine variability events are expected to be of the order of $\sim
10$ days, significantly shorter variations (day long or even down to a
few hours) are seen (e.g. PKS 1510-089 \cite{marscher10a}). These can
be explained if the observed variations come from a fraction of the
jet cross section.

Recently, a diagnostic has been proposed \cite{dotson11} for the location of the 
blazar GeV emission that is based on the energy dependence of  Fermi flux variations. 
This diagnostic makes use of the fact that if the blazar emission takes place inside the BRL the cooling is done on the $\sim 10 $ eV  BLR photons, while if it takes place outside the BLR, the most abundant seed photons are
IR photons ($\sim 0.1 $ eV of the MT.  In the first case cooling
 takes place at the onset of the Klein-Nishina cross section and GeV
 variability is  achromatic, while in the second case cooling takes place in the Thomson regime and variability is faster at higher energies.
 Although this is a powerful diagnostic, it can only
be applied to a small number of  bright {\sl Fermi} flares.

\section {\bf An SED-based {\boldmath $\gamma$}-ray location diagnostic and
  its  application}  In powerful blazars, the emission consists of two
spectral components. The low frequency one peaks at $\nu_s \sim
10^{13}$ Hz \cite{giommi11} and is attributed to synchrotron
radiation; the high frequency one peaks at $\nu_c\sim 10^{22}$ Hz
\cite{giommi11} and is usually attributed to inverse Compton emission
from external photons that are produced in the broad line region or in
the MT.  
% The possibility of synchrotron-self Compton (SSC) is disfavored, as modeling such sources {\sl self-consistently} in the SSC framework results unavoidably \cite{georganopoulos12} in pc-scale source sizes, incompatible with the few hour variability times observed, even after relativistic beaming has been taken into account.

%Returning to the EC framework,
 If $\Gamma$ is the bulk Lorentz factor
and $\delta$ the usual Doppler factor of the jet flow, and $\gamma_b$
is the Lorentz factor of the electrons responsible for the synchrotron
and EC SED peaks, then,
 %assuming $\delta=\Gamma$, as anticipated for sources  with jets well aligned to the line of sight, then
\begin{equation}
\epsilon_s={B\over B_{cr} }\gamma_b^2 \delta, 
\end{equation}
\begin{equation}
\epsilon_c={4 \over 3} \epsilon_0\gamma_b^2\delta^2,
\label{eq:ec}
\end{equation}
where $\epsilon_s$ is the synchrotron peak energy, $\epsilon_c$ is the
EC peak energy, $\epsilon_0$ is the characteristic energy of the
external photon field, all in units of the electron rest mass, $B$ is
the magnetic field permeating the emission region and $B_{cr}=(m_e^2
c^3)/( e \hbar)=4.4 \times 10^{13}$ G is the critical magnetic
field. Note that equation (\ref{eq:ec}) is valid only if the
scattering of electrons with Lorentz factor $\gamma_b$ takes place in
the Thomson regime.  The condition for this, $\epsilon_0\gamma_b
\delta \lesssim 1$ can be written with the help of equation
(\ref{eq:ec}) as $( \epsilon_c \epsilon_0)^{1/2}\lesssim1$.  The
highest possible energy external seed photons are UV line photons with
$\epsilon_0 \approx 10^{-4}$, which means that the scattering is
indeed in the Thomson regime as long as $\epsilon_c\lesssim
\epsilon_0^{-1}\approx 10^4$. This corresponds to an energy of
$\approx 5$ GeV or $\nu_c \approx 10^{24}$ Hz. Given that in most cases powerful
blazars peak at lower $\nu_c$, with $<\nu_c>\approx
10^{22}$ Hz \cite{giommi11}, the scattering of the electrons producing
the EC peak is well within the Thomson regime.  Taking the ratio of
the two peak energies we obtain:
\begin{equation}
{B\over \delta}={4 \epsilon_0 \epsilon_s B_{cr} \over 3  \epsilon_c}
\label{eq:eratio}
\end{equation}
The same ratio $B/\delta$ can be obtained from the expression for the
Compton dominance $k$, the ratio $L_c/L_s$ of EC to synchrotron
luminosity:
\begin{equation}
k={L_{c}\over L_{s} }={U_0' (\delta^6/\Gamma^2) \over U_B \delta^4 }={32\pi   \delta^2 U_0  \over  3 B^2},
\label{eq:k}
\end{equation} 
where $U_0$ is the external photon field energy density in the galaxy
frame, $U_0'=(4/3)U_0 \Gamma^2$ is the external photon field energy
density in the jet comoving frame and $U_B=B^2/(8\pi)$ is the magnetic
field energy density \cite{dermer95, georganopoulos01}.  Solving
equation (\ref{eq:k}) for $B/\delta$ and equating to equation
(\ref{eq:eratio}), we obtain our final expression
\begin{equation}
{ U_0^{1/2} \over \epsilon_0}=\sqrt{{ k B_{cr}^2 \over 6\pi }} { \epsilon_s \over \epsilon_c}=3.2 \times 10^{4}\, { k_1^{1/2}\, \nu_{s,13} \over \nu_{c,22}} \; {\rm G}
\label{eq:diag}
\end{equation}
where $k_1$ is the Compton dominance in units of 10, $\nu_{c,22}$ is
$\nu_c$ in units of $10^{22}$ Hz, and $\nu_{s,13}$ is $\nu_s$ in units
of $10^{13}$ Hz. {\sl Note that the RHS  contains only
  observables. It is the RHS  that informs
  us about the ratio of the square root of the energy density over the
  peak energy of the seed photons (seed factor, SF, in Gauss) available at the
  location of the $\gamma$-ray emission.It is this information that we
  can use for understanding where the emission comes from.}

\subsection{ The SF in the BLR and in the MT.}

  Reverberation mapping finds
that $R_{BLR} \approx 1-3 \times 10^{17} L_{d,45}^{1/2}$ cm where
$L_{d,45}$ is the accretion disk luminosity $L_d$ in units of
$10^{45}$ erg s$^{-1}$, and that a fraction $\xi\sim 0.1$ of $L_d$ is
reprocessed by the BLR \cite{kaspi07,bentz09}.  The energy density for
the BLR is then $U_{0} =\xi L_{d}/(4\pi R^2 c)\approx0.29- 2.6\times
10^{-2}$ erg cm$^{-3}$ \cite{ghisellini09}. Because $R\propto
L_d^{1/2}$, $U_{0}$ is the same for sources of different luminosities.
The BLR SED in the galaxy frame can be approximated by a blackbody
peaking at $\nu_{0} = 1.5 \nu_{\textrm{Ly}\alpha}$ ($\epsilon_{0}
\approx 3 \times 10^{-5}$, \cite{tavecchio08}).  Using these we obtain
$SF \sim 1.8-5.5 \times 10^{3}$ G.

Because of the larger distance of the MT from the central engine,
reverberation mapping has only been performed for Seyferts
(e.g. \cite{suganuma06}), lower luminosity and therefore smaller MT
size sources.  These studies, along with IR interferometric studies
(e.g. \cite{kishimoto11}) are also in agreement with an $R\propto
L_d^{1/2}$ scaling, suggesting that $U_0$ is the same for sources of
different luminosities.  Adopting the results of \cite{malmrose11} for
the blazar 4C 21.35, a blackbody of temperature $T =1200$ K ($\epsilon_0
= 5.7 \times 10^{-7}$) and $L=7.9 \times 10^{45}$ erg s$^{-1}$ emitted
from a radius of $\sim 1-2$ pc, we obtain $U_0 = 0.5-2.2 \times
10^{-3} \textrm{ergs cm}^{-3}$.  With these we obtain $SF \sim 4.1-8.2
\times 10^{4}$ G. A comparison of these indicative ranges
is shown in Figures 2 and 3.

\begin{figure}[t]
\centerline{
\includegraphics [width=3.1in]{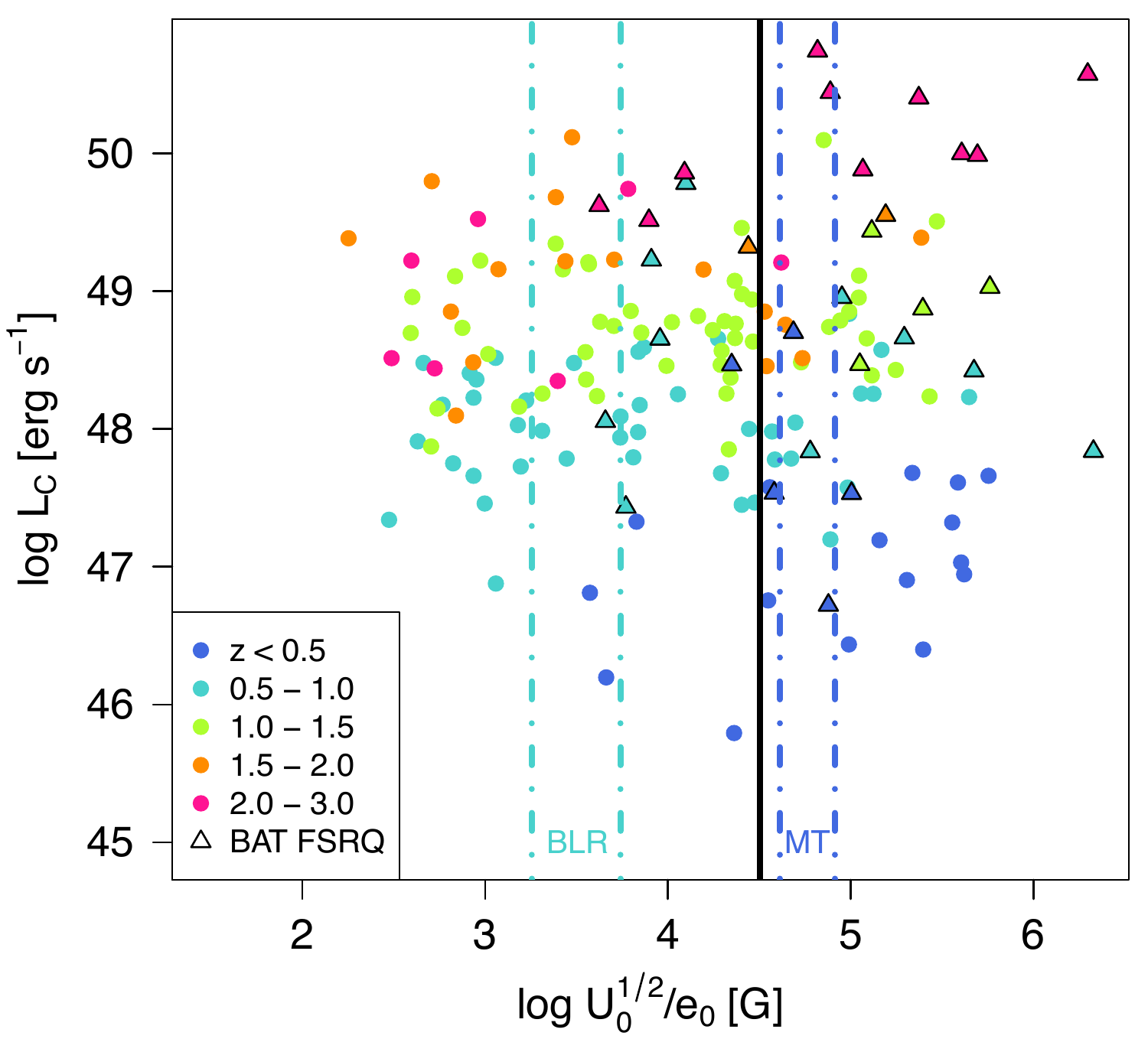}}
\caption{   $L_c$ as a function of the SF for a number of
  sources, color-coded for redshift, for which $\nu_s$, $\nu_c$ and
  $k$ are found from their SEDs. The solid line is the average from
  \cite{giommi11}, the areas marked by dotted lines correspond to the
  $\gamma$-ray emission site being inside the BLR (left) or the MT
  (right).}
\label{eileen}
\end{figure}

\subsection{Can  SSC be dominant?}

In the previous discussion we considered only external photons.
Here we briefly discuss under what conditions synchrotron photons
would be the dominant seed photons.
A lower limit for he synchrotron photon density $U_s$ in the comoving frame (assuming $\delta=\Gamma$
is 
\begin{equation}
U_s={L_s \over 4 \pi c^3 t_{var}^2 \Gamma^6},
\end{equation}
where $t_{var}$  is the observed variability timescale  and $L_s$ the luminosity of the synchrotron component. For this to dominate over $U_0 \Gamma^2$, the comoving external photon field energy density, we require 
\begin{equation}
\Gamma < \left({L_s \over 4 \pi c^3 t_{var}^2 U_0}\right)^{1\over 8}=12.6 \left({L_{s,47} \over t_{var,6h} U_{0,-4}}\right)^{1\over 8}.
\end{equation}
For  our adopted BLR range,
$U_{0} \approx 0.29- 2.6\times 10^{-2}$ erg cm$^{-3}$, this is equivalent to
$\Gamma <8.4-11.0 \; L_{s,47}^{1/8} t_{var,6h}^{-1/8}$
Similarly for our adopted MT range, 
$U_{0} = 0.5-2.2 \times
10^{-3} \textrm{erg cm}^{-3}$ this is equivalent to
$\Gamma <11.4-13.7 \; L_{s,47}^{1/8} t_{var,6h}^{-1/8}$.
VLBI studies of superluminal speeds in FSRQs \citep[e.g. see figure 24 of ][]{jorstad05} show that for most FSRQs, $ \Gamma \gtrsim 10$, with values reaching up to $\sim 40$.
This raises the possibility that our diagnostic is  relevant for sources that have relatively high $\Gamma$, and that for the slower sources SSC may dominate.

\section{ Preliminary Results.} 
 
Some preliminary results using the sample of \cite{meyer11} with
detections in the 2FGL or BAT are shown in Figures 2 and 3. The
average values of $k\sim$few, $\nu_s=10^{13}$ Hz, and $\nu_c=10^{22}$
found for powerful blazars \cite{giommi11} correspond to a SF (solid line)
  that is closer to the MT range than that of the BLR.  The fact that
the two  bands corresponding to the SF we estimated for the BLR
and the MT occupy the central part of the observed SF range is 
encouraging and indicates that FSRQs experience a seed photon
environment that is  within the range of what is expected from our
simple BLR and MT considerations. We note that estimating $\nu_s$ and
$\nu_c$ requires good multi-wavelength coverage, and errors of a
factor up to $\sim 10$ are possible, particularly in the current
estimates of $\nu_c$. This suggests that the actual range of SF is
significantly more narrow than the current data suggest, likely giving
closer agreement with the range denoted by the bands in Figures 2 and
3 once we have completed our SED analysis on very well sampled
sources.

  In future work, we plan to use all available data,
including archival data from sources like NED with newer catalogues
from WISE, Planck, SWIFT, BAT, LAT to produce SEDs with simultaneous
or contemporaneous data for as many sources as possible and derive
their $\nu_s$, $\nu_c$ and $k$ following \cite{meyer11}, along with
error estimates for these quantities. We anticipate that this work will significantly
expand the sample in \cite{meyer11} which was restricted to coverage
of the synchrotron component only.
 
To estimate the actual range of the SF in the BLR and the MT,
we need to know how the quantity $U_0^{1/2}/\epsilon_0$ is expected to
vary radially. This requires information on the BLR and the MT
stratification, something that is not yet well constrained
\cite{kaspi07,bentz09,landt10}) yet. For the modeling component of
the project, we plan to explore how the SF range widens if one considers
the stratification of different line photons in the BLR
(e.g. \cite{poutanen10} and the fact that powerful IR components with
$T$ down to $\sim 300$ K have been detected in radio-loud AGN \cite{landt10}.
 
 \begin{figure}[t]
\centerline{ \includegraphics [width=3.1in]{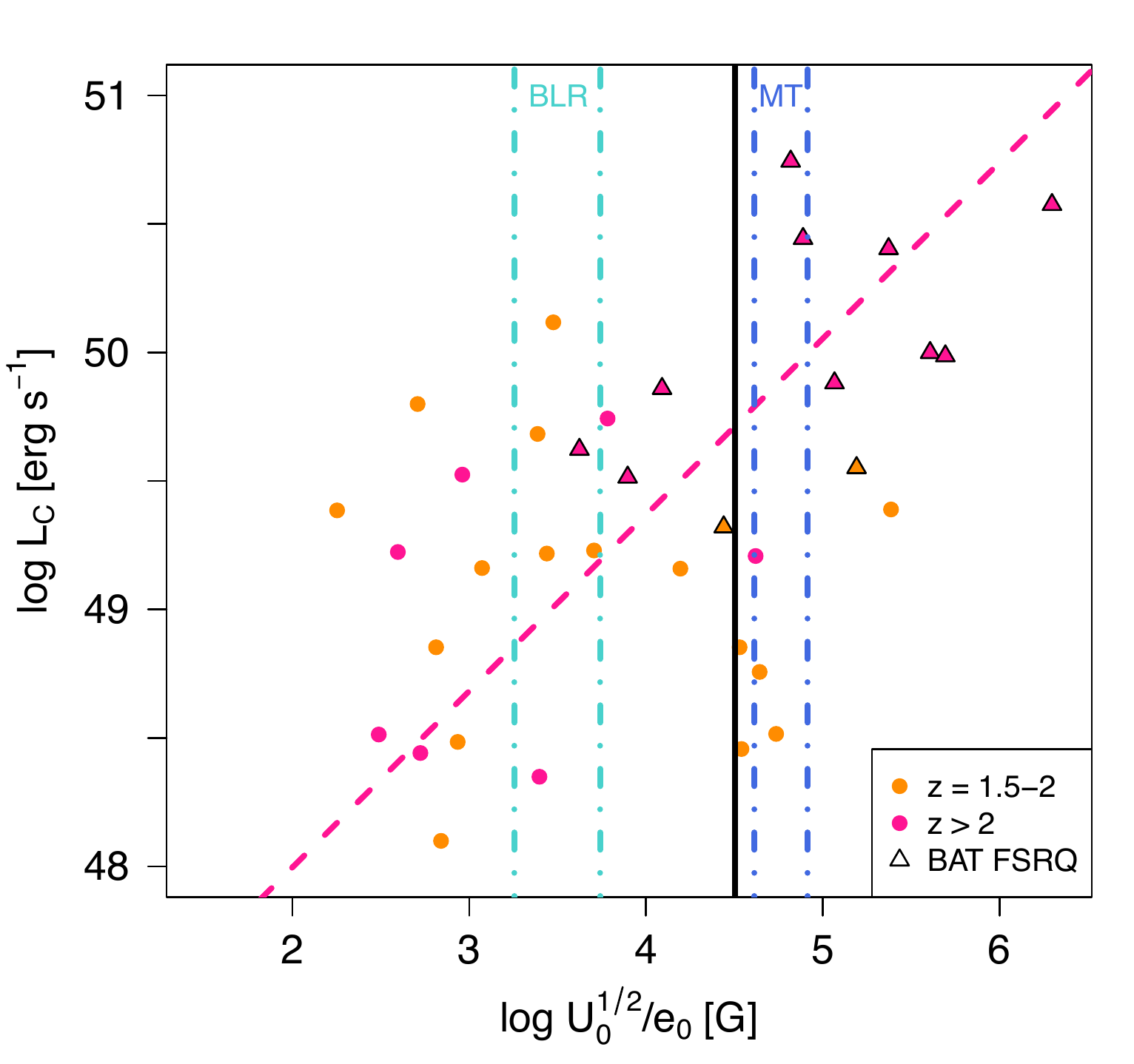}}
\caption{ \sl \small Same as Figure 2, but for sources with
  $z>2$. There is a clear correlation between the total IC power and
  the seed factor SF, moving from values consistent with the BLR to
  the MT at highest energies. \vspace{-10pt}}
\label{eileen2}
\end{figure}

%Kryfo sxoleio:

%We will also examine trends like the one suggested in Figure 3: for high-$z$ sources, the SF shifts from the BLR to the MT as $L_c$ increases. If this trend is verified, it will suggest that the location of emission moves outward as the source power increases. This could be explained by taking into account that more powerful jets have higher $\Gamma$ \cite{kharb10} and considering that the location of the emission scales with $\Gamma^2$, as in the internal shock model \cite{spada01}.  This could fit into a picture where lower $\Gamma$ sources are SSC dominated  and only  for higher $\Gamma$ the  emission becomes EC dominated.

\end{document}